\documentclass[twocolumn, a4paper, 10pt]{article}
\usepackage[utf8]{inputenc}
\usepackage[T1]{fontenc}
\usepackage{secdot}
\usepackage{graphicx}
\usepackage{color}
\usepackage{sectsty}
\usepackage[usenames,dvipsnames,svgnames,table]{xcolor}
\usepackage[a4paper, hmargin=2cm, tmargin=1.5cm, bmargin=3cm, footskip=2cm]{geometry}
\usepackage{fancyhdr}
\begin{document}

\sectionfont{\centering\large}
\twocolumn[
\begin{@twocolumnfalse}
\begin{center}
\title{Zeeman optical pumping of $^{87}$Rb atoms in a hollow core photonic crystal fibre}
\date{}
\maketitle
\end{center}
\textbf{\author{Tomasz~Krehlik$^{1,2,*}$, Artur~Stabrawa$^{2}$, Rafał~Gartman$^{2}$, Krzysztof~T.~Kaczmarek$^{2}$, Robert~L\"ow$^{3}$, and Adam~Wojciechowski$^{1}$}}
\\ \textit{$^1$ Photonics Department, Jagiellonian University, Łojasiewicza 11, 30-348 Kraków, Poland}
\\ \textit{$^2$ ORCA Computing Ltd, London, UK}
\\ \textit{$^3$ University of Stuttgart, 5. Institute of Physics and Center for Integrated Quantum Science and Technology IQST, Pfaffenwaldring 57, 70569 Stuttgart, Germany}
\\ \textit{$^*$ tomasz.krehlik@doctoral.uj.edu.pl}

\begin{quote}\small{
\begin{quote}
\begin{center}
\textbf{Abstract}
\end{center}
Preparation of an atomic ensemble in a particular Zeeman state is a critical step of many protocols for implementing quantum sensors and quantum memories. These devices can also benefit from optical fibre integration. In this work we describe experimental results supported by a theoretical model of single-beam optical pumping of $^{87}$Rb atoms within a hollow-core photonic crystal fibre. The observed 50$\%$ population increase in the pumped F=2,~m$_F$=2 Zeeman substate along with the depopulation of remaining Zeeman substates enabled us to achieve a 3 times improvement in the relative population of the m$_F$=2 substate within the F=2 manifold, with 60$\%$ of the F=2 population residing in the m$_F$=2 dark sublevel. Based on our theoretical model, we also propose methods to further improve the pumping efficiency in alkali-filled hollow-core fibres.
\end{quote}
}
\end{quote}
 \bigskip
    \bigskip
\end{@twocolumnfalse}
]

\section{Introduction}
With the advent of new and reliable manufacturing techniques and novel designs of microstructurised optical fibres \cite{bib:russel} a specific class of hollow core photonic crystal fibres (HCPCF) is gaining in prominence in the fields of quantum optics \cite{bib:sprague}, gas sensing \cite{bib:nikodem, bib:belardi}, high-power pulses delivery \cite{bib:michieletto} and telecommunications \cite{bib:zhao, bib:sakr}. This uptake of new fibre types arises from technical advantages that those fibres bring, namely low to no group velocity dispersion \cite{bib:kolyadin}, high tunability of mode size \cite{bib:russel}, high damage threshold \cite{bib:michieletto}, wideband operation \cite{bib:belardi}, and unique ability to introduce optically active materials (e.g. atomic vapours) within the light-guiding region. One of the most interesting and promising applications for atom-filled HCPCF is as a platform for realisation of single photon delay lines and quantum memories, which serve for short-term operation buffering in quantum computing \cite{bib:lvovsky} and quantum communications \cite{bib:rep}. A possible implementation is the off-resonant cascaded absorption (ORCA) quantum memory protocol \cite{bib:kris}, relying on the coherent interaction of alkali metal atoms in a gas phase with two light fields. Filling a HCPCF with the atomic vapour and coupling both light beams into its core will allow for maintaining a perfect beams' overlap and high coupling beam intensity on a long path, which is beneficial for memory efficiency. This would require much more power in a free space because of the diffraction enforcing a compromise between the beam’s waist and divergence. Simultaneously, for HCPCFs with a core diameter of tens of $\mu$m, direct splicing with solid-core fibres is possible \cite{bib:stawska, bib:thapa}, enabling mechanically robust light delivery and easy all-fibre integration of HCPCF-based memories into larger systems, which is needed for an industrial application.

In this work, we focus on the optical pumping of rubidium atoms confined in the core of a kagome-type fibre \cite{bib:robert} to the F=2,~m$_F$=2 spin state (stretched state), which improves the efficiency of the ORCA memory. In the case of an unpolarised medium, the two-photon 5S$_{1/2} \rightarrow$ 5D$_{5/2}$ transition exploited in the ORCA scheme can be realised through various hyperfine levels of the intermediate 5P$_{3/2}$ state, and the interference of probability amplitudes between these excitation paths produces a beat-note characteristic of memory efficiency over time \cite{bib:kris}. On the other hand, when the medium is pumped to the stretched state and interacts with $\sigma_+$ polarised beams, only one intermediate state is involved, which eliminates the interference and therefore maximises the efficiency over the entire lifetime. The influence of state preparation on memory efficiency is analysed and experimentally demonstrated in Ref. \cite{bib:finkelstein} for the case of a standard spectroscopic cell.

Aiming at practical applications, we investigate the efficiency of Zeeman pumping in a most basic scheme, i.e. without applying a magnetic field and using only one $\sigma_+$ polarised pump beam (without the repumping light). The D1 transition is chosen for pumping light in order to minimise its influence on the memory excited levels (5P$_{3/2}$ and 5D$_{5/2}$) and to enable its efficient separation from the signal photons at D2 line in the ORCA memory scheme \cite{bib:kris}. In this pumping scheme two kinds of dark states exist - the F=2,~m$_F$=2 state (because of the selection rules for the light polarisation used), and the F=1 manifold (because of a large detuning of the pump beam). The relation between efficiencies of pumping to these two dark states depends on specific values of the effective pump Rabi frequency and its polarisation purity, which in combination with a quick transit decay (on the order of natural decay for our HCPCF diameter) calls for experimental examination of the pumping process. The pumping efficiency was estimated through the measurements of absorption spectra of a weak resonant probe beam and comparison with spectra calculated from the steady-state solution of the Lindblad master equation \cite{bib:lindblad} taking into account all Zeeman sublevels involved in the interaction with pump and probe beams. The model also includes transient effects from atoms leaving and entering the probe volume by additional decay and repopulation.

\begin{figure}[htpb]
\centering
\fbox{\includegraphics[width=0.95\linewidth]{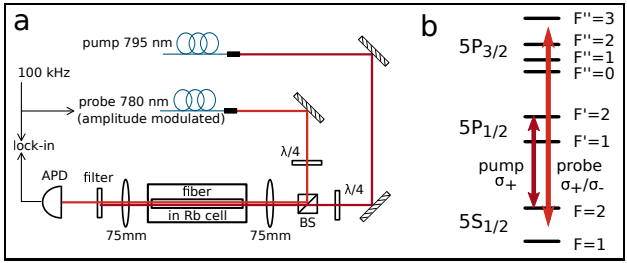}}
\caption{a) Simplified experimental setup. The pump and probe beams are delivered by PM fibres, overlapped on a free-space non-polarising beamsplitter and coupled (copropagating) into the HCPCF filled with Rb vapour, which is placed in an oven. Beams polarisations are controlled with quarter-wave plates. After passing through the fibre the pump beam is blocked with an interference filter, while transmission of the probe beam is measured with an avalanche photodiode. b) Scheme of $^{87}$Rb levels used. The pump beam is $\sigma_+$ polarised and tuned to resonance with F=2 – F’=2 transition (D1 line), while probe is either $\sigma_+$ or $\sigma_-$ polarised and scanned through the entire D2 line.}
\label{fig:setup}
\end{figure}

\section{Experimental details}

The key elements of the experimental setup are presented in Fig \ref{fig:setup}a. Its central part is a 7.5-cm-long glass spectroscopic cell filled with natural abundance rubidium vapour. Inside the cell, a glass capillary is suspended which holds the kagome-type optical fibre with a core diameter of around 60~$\mu$m. Details of this novel design of a fibre enclosed in a glass cell have been recently described in Ref. \cite{bib:robert}. Exposed ends of the fibre which is approximately 1~mm shorter than the cell's inner length, allow for its self-filling with Rb atoms on the timescale of days. The cell is kept in a temperature-stabilised oven, with a temperature tuned between 40 and 120~°C to regulate the atomic density inside the cell and to speed up the initial fibre filling process. Two laser beams, pump and probe, are overlapped on a non-polarising beamsplitter and coupled into the fibre using a single lens (75~mm focal length). The polarisation of the beams is set independently using quarter-wave plates. The stronger (pump) beam, with the power varied between 1~$\mu$W and 20~mW (corresponding to 10$^2$ - 10$^6$ saturation intensities, calculated for 30~$\mu$m mode field diameter which is the lower estimate), has a fixed $\sigma_+$ circular polarisation, and is tuned to resonance with the 5S$_{1/2}$,~F=2 $\rightarrow$ 5P$_{1/2}$,~F’=2 transition, as shown in Fig. \ref{fig:setup}b. The weaker (probe) beam, with power of the order of few nW (corresponding to 0.1 saturation intensity), has its polarisation changed between $\sigma_+$ and $\sigma_-$ . Its frequency is swept through the entire spectral range of the 5S$_{1/2}$ $\rightarrow$ 5P$_{3/2}$ transition, and its transmission through the vapour-filled fibre is recorded with an avalanche photodiode working in the linear (intensity) mode. Probe beam’s power is chosen to be well below the saturation intensity, firstly not to introduce any perturbations to the medium, secondly to limit the number of the theoretical model’s parameters. In order to discriminate the transmitted probe light from the strong pump background, an interference filter is placed in front of the photodiode. To further improve SNR, the probe beam is amplitude-modulated with an AOM at 100~kHz, and the photodiode output is demodulated by a lock-in amplifier.

During the first two weeks of the experiment, the cell was baked at 120~°C (with its stem kept a few degrees colder) to facilitate Rb atoms penetration along the whole length of the fibre core. After this period we noticed no further change in atomic concentration inside the fibre, which is in agreement with the rapid fibre filling on timescales of few days reported in Ref. \cite{bib:robert}. The on-resonance (F=2 $\rightarrow$ F"=1,2,3 transition) optical depth (OD) estimated from a fit to a probe transmission spectrum was above 50, which means that the fiber's core filling with rubidium atoms was sufficient for a future implementation of the ORCA memory protocol. However, such OD turned out to be too high to conduct quantitative optical pumping measurements (because of absorption profiles being saturated over frequency regions as wide as 2~GHz) and after the bake-out period the fibre temperature was decreased to 40~°C.

\begin{figure}[h!]
\centering
\fbox{\includegraphics[width=0.8\linewidth]{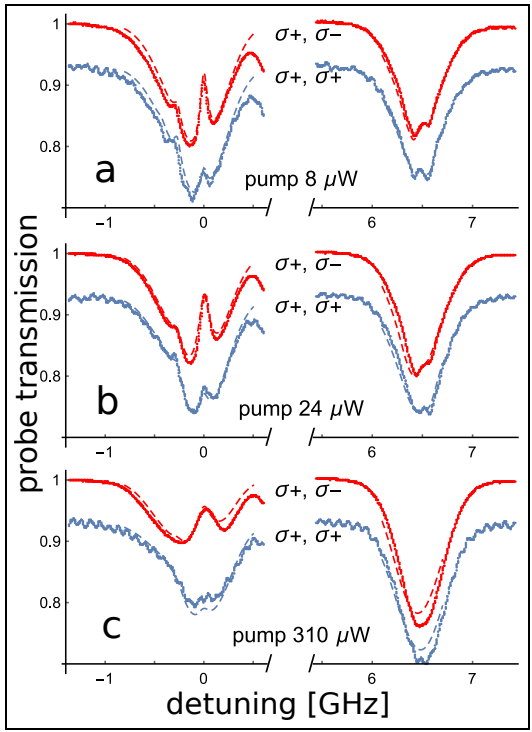}}
\caption{Comparison of the measured probe transmission spectra (thick) and simulations (dashed) for different pump powers. In each plot the upper trace (labeled $\sigma_+$, $\sigma_-$) corresponds to the opposite circular polarisations of the pump and probe beams, while the lower one (artificially moved downwards) to the same circular polarisation. The left absorption spectrum corresponds to transitions from the F=2 groundstate of $^{87}$Rb and the right one - from the F=1 (absorption lines of  $^{85}$Rb are not shown). The narrow features of increased transmission result from both velocity-selective optical pumping and EIT. Pump Rabi frequencies used in the simulations are 450~MHz, 800~MHz and 2800~MHz for plots a, b and c respectively.}
\label{fig:spectra}
\end{figure}

The optical pumping measurement was conducted as follows. The beams were continuously switched on. The pumping beam was tuned to resonance (F=2 $\rightarrow$ F'=2, D1 line) and its power was set to the desired value. Then the polarisation of the probe was set to have the same handiness as the pump, and its frequency was swept through the D2 line while the transmission signal was recorded. Subsequently, the probe polarisation was switched to the opposite handiness, and the spectrum acquisition was repeated. The resultant spectra were then normalised and analysed. To frequency-calibrate the scans, an independent reference saturated-absorption spectroscopy was performed in a standard vapour cell.

\begin{figure}[htpb]
\centering
\fbox{\includegraphics[width=0.95\linewidth]{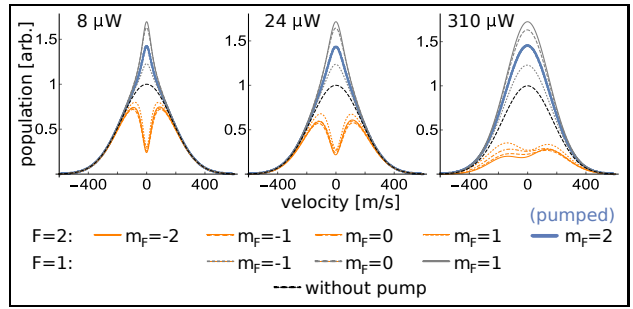}}
\caption{Velocity distributions simulated for the same parameters as in Fig. \ref{fig:spectra}. Thick blue line corresponds to the F=2,~m$_F$=2 state, orange lines to other F=2 sublevels, gray to F=1 sublevels. The black dashed line is a reference thermal distribution for no pump present. On resonance ($v$=0) the maximum pumping efficiency is obtained already for the weakest pump (8~$\mu$W); increasing pump power leads to widening of the efficiently pumped region (via power broadening). The asymmetry seen for 310~$\mu$W results from pumping through the off-resonant F’=1 state, which is more efficient for negative-velocity atoms (Doppler-shifted towards the resonance).}
\label{fig:distributions}
\end{figure}

\section{Theoretical model}
In order to have a good understanding of the optical pumping process in Rb atoms confined within a relatively narrow space of the HCPCF and to get quantitative information on the pumping process efficiency, we performed a theoretical modelling of the recorded spectra. The developed model that was compared to the measured spectra was a 32-level atom (including all Zeeman sublevels of 5S$_{1/2}$, 5P$_{1/2}$ and 5P$_{3/2}$ states), for which a steady-state density matrix was calculated from the Lindblad master equation \cite{bib:lindblad}, taking into account natural and transit decays and interactions with a $\sigma_+$ polarised pump of various intensities and a weak, $\sigma_+$ or $\sigma_-$ polarised probe. To solve the model, dipole and rotating wave approximations were made. Doppler broadening was accounted for by solving the model independently for different velocity classes of the thermal distribution of atoms. A simple 3D Maxwell-Boltzmann distribution of velocities was assumed, neglecting any possible correlations after atom collisions with fibre walls.  Then the theoretical transmission spectrum for the probe was calculated according to the formula \cite{bib:lindblad}:

\begin{equation}
\alpha(f) \sim \int_v dv  p(v) \sum_{g,e} Im(\rho_{ge}(f,v) \times d_{ge}),
\end{equation}

\noindent where $\alpha$ is absorption coefficient, $f$ is nominal probe detuning, $p(v)$ is the velocity probability distribution, $g$ and $e$ enumerate ground and excited states of the probe transition respectively, $\rho_{ge}$ is the corresponding density matrix element (which is frequency and, through the Doppler effect, velocity dependent) and $d_{ge}$ the respective dipole moment.
The proportionality coefficient above includes atomic constants and the atoms' concentration; its value was inferred from the experiment and set such that spectra simulated for an unpumped ensemble matched the measurement. Other parameters were as follows: spontaneous relaxation rate $2\pi \cdot 6$~MHz, probe laser width $2\pi \cdot 0.3$~MHz, pump laser width $2\pi \cdot 6$~MHz, transit relaxation rate $2\pi \cdot 1.8$~MHz, vapour's temperature 40~°C The pump beam’s Rabi frequency used for modelling was calculated for each measurement from the experimental value of pump power, but with the scaling factor of ¼ (common for all powers), which improved the agreement with measurements. This scaling factor may be attributed to averaging over the pump intensity distribution within the optical mode guided by the fibre structure.

\section{Results}

\begin{figure}[htpb]
\centering
\fbox{\includegraphics[width=0.8\linewidth]{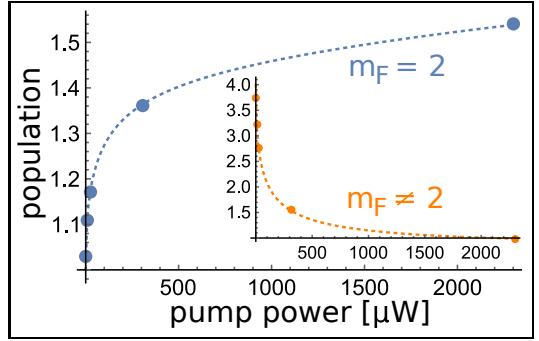}}
\caption{Dependence of the calculated pumping efficiency on pump power. Circles correspond to powers for which the simulations were verified by comparison with the measured spectra. The blue (main) plot is population in F=2,~m$_F$=2 state (integrated over the whole velocity distribution). Orange plot (inset) is the sum of populations in all other Zeeman substates of the F=2 state (inset’s axes in the same units as for the main plot). The populations are expressed in units of the equilibrium (thermal) population in a single Zeeman substate.}
\label{fig:efficiency}
\end{figure}

Comparison between the measured and simulated spectra is shown in Fig. \ref{fig:spectra}. A good agreement for both polarisations allows us to assume that our model properly describes the pumping process. The population distributions calculated from the model are shown in Fig. \ref{fig:distributions}, and they serve as a basis for the pumping efficiency estimation. The population of the F=2,~m$_F$=2 (stretched) state integrated over the whole velocity distribution is plotted in Fig. \ref{fig:efficiency} as a function of the pump power. The results provide information on the single-beam optical pumping efficiency attainable in our experiment. The pumping efficiency begins to reach its maximum for several mW, with the population increase in the stretched state of more than 50$\%$. Applying even stronger pump beam (tens of mW) resulted in noticeable AC Stark-shifts and this regime is not analysed here. Simultaneously with increasing the stretched state's population, the total occupation of other Zeeman levels of the F=2 state decreases by the factor of 4. Hence the polarisation of the F=2 state (being one of critical parameters influencing quantum memory efficiency and defined here as the ratio of the population in the stretched state to the population of the whole F=2 manifold) improves from 20$\%$ (without pumping) to 60$\%$.

As seen from Fig. \ref{fig:distributions}, a significant proportion of atoms is also pumped to the F=1 groundstate. While this is not detrimental for memory operation (because of a far detuning of this state from the memory transitions), it limits the obtainable atomic density in the F=2,~m$_F$=2 state. This indicates that a repumping beam would be necessary for attaining higher pumping efficiencies. Preliminary calculations based on rate equations let us expect additional threefold increase in a stretched state population after incorporating a repumping beam at F=1 $\rightarrow$ F'=2 transition. Another conclusion from Fig. \ref{fig:distributions} is that the on-resonance ($v$=0) population of the stretched state saturates already for the weakest pump shown. Increasing pump power leads to power-broadening of the transition, resulting in widening of the efficiently pumped velocity range. This suggests that pumping with spectrally broader light could require less power, which would be profitable for industrial scalability (especially if fibres with wider cores are used). It will also facilitate adding the repumping beam. Spectral broadening suppresses coherent effects like EIT, which should be considered in the case of two strong, spectrally narrow fields of the pumping and repumping beams, sharing the common F'=2 state.

\begin{figure}[htpb]
\centering
\fbox{\includegraphics[width=0.8\linewidth]{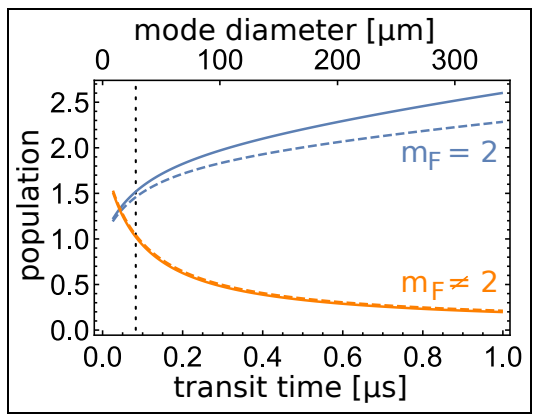}}
\caption{Calculated pumping efficiency dependence on the mode field diameter for purely $\sigma_+$ polarised pump (solid lines) and for a 5$\%$ admixture of $\sigma_-$ polarisation (dashed lines). The upper (blue) lines correspond to the F=2,~m$_F$=2 state, while the bottom ones (almost overlapping) to the sum in other F=2 states. The pump Rabi frequency is 7700~MHz. Populations are presented in units of the equilibrium Zeeman substate population. The dotted vertical line indicates the approximate parameters of our experiment.}
\label{fig:transit}
\end{figure}

Another factor limiting the pumping efficiency in our set-up is a short transit time (and therefore high relaxation rate) resulting from the hollow-core diameter. Calculated dependence of the pumping efficiency on transit time is presented in Fig. \ref{fig:transit}, along with a mode diameter resulting in a given transit time (assuming the cell temperature of 40~°C). Shifting to a fibre with a mode field diameter of 300~$\mu$m results in 150$\%$ population increase in F=2,~m$_F$=2 state. Combined with enhanced depopulation of the other Zeeman substates, this leads to the polarisation of the F=2 manifold of 90$\%$. These results don't get significantly worse even if a 5$\%$ admixture of $\sigma_-$ polarisation is present in a pumping beam (dashed lines in Fig. \ref{fig:transit}), and circular polarisation was checked to be maintained in our fibre with the purity of above 99$\%$, which allows us to treat the obtained numbers as a realistic estimate. Currently the difficulty in using such wide fibres is their multimodal operation, but there is a work ongoing which aims at designing structures suppressing higher order modes propagation \cite{bib:zhang, bib:habib}. Techniques of efficient light coupling from solid-core fibres to HCPCFs with large diameter mismatch are being improved \cite{bib:chen}. The benefit from using even wider fibres is moderate, and it would be technically challenging.

\section{Conclusion}
The efficiency of optical pumping to the F=2,~m$_F$=2 Zeeman substate was investigated for vapours of $^{87}$Rb contained in a HCPCF. The results should also be directly applicable to other types of hollow-core fibres. With only one, $\sigma_+$ polarised pumping beam operating at D1 line, above 50$\%$ increase in the stretched state population (translating to 60$\%$ polarisation of the F=2 manifold) was measured already for 2~mW pump power. Three main paths of improving the pumping efficiency were identified: applying the repumping beam (to prevent atoms' escape to the F=1 groundstate), using a fibre with a wider core (to lower the transit relaxation rate) and spectrally broadening the pump (to obtain a wide velocity range of efficient pumping in a more power-effective way than by relying on power-broadening of the pump beam).

\bigskip
\textbf{Funding.} This work was partially funded through the MNS support program of the Faculty of Physics, Astronomy and Applied Computer Science at the Jagiellonian University in Krakow. This work was also supported by U.S. AFOSR EOARD FA8655-21-1-7059.\\

\textbf{Acknowledgments.} We acknowledge Nicolas Joly from Max Planck Institute for Light, Erlangen for providing the fibre and Daniel Weller and Frank Schreiber from University of Stuttgart for making the cell. We are thankful to Josh Nunn for discussions.\\

\textbf{Disclosures.} The authors declare that there are no conflicts of interest related to this article.\\

\textbf{Data availability.} Data underlying the results presented in this paper are not publicly available at this time but may be obtained from the authors upon reasonable request.


\begin{thebibliography}{19}
\bibitem{bib:russel}
P. S. J. Russel,
  Photonic-Crystal fibers, Journal of Lightwave Technology \textbf{24},
  4729 (2006).
\bibitem{bib:sprague}
M. Sprague et al.,
 Broadband single-photon-level memory in a hollow-core photonic crystal fiber, Nature Photonics \textbf{8},
  287 (2014).
\bibitem{bib:nikodem}
M. Nikodem et al.,
  Demonstration of mid-infrared gas sensing using an anti-resonant hollow core fiber and a quantum cascade laser, Optics Express \textbf{27},
  36350 (2019).
\bibitem{bib:belardi}
W. Belardi,
  Design and Properties of Hollow Antiresonant fibers for the Visible and Near Infrared Spectral Range, Journal of Lightwave Technology \textbf{33},
  4497 (2015).
\bibitem{bib:michieletto}
M. Michieletto et al.,
 Hollow-core fibers for high power pulse
delivery, Optics Express \textbf{24},
  7103 (2016).
\bibitem{bib:zhao}
X. Zhao et al.,
 5-tube hollow-core anti-resonant fiber with ultralow loss and single mode, Optics Communications \textbf{501},
 127347 (2021).
\bibitem{bib:sakr}
H. Sakr et al.,
 Hollow core optical fibers with comparable attenuation to silica fibers between 600 and 1100~nm, Nature Communications \textbf{11},
  6030 (2020).
\bibitem{bib:kolyadin}
A. N. Kolyadin et al.,
  Negative curvature hollow-core fibers: dispersion properties and
femtosecond pulse delivery, Physics Procedia \textbf{73},
  59 (2015).
\bibitem{bib:lvovsky}
A. I. Lvovsky et al.,
  Optical quantum memory, Nature Photonics \textbf{3},
  706 (2009).
\bibitem{bib:rep}
N. Sangouard et al.,
  Quantum repeaters based on atomic ensembles and linear optics, Reviews of Modern Physics \textbf{83},
  33 (2011).
\bibitem{bib:kris}
K. T. Kaczmarek et al.,
 High-speed noise-free optical quantum memory, Physical Review A \textbf{97},
  42316 (2018).
\bibitem{bib:stawska}
H. Stawska et al.,
 Combining Hollow Core Photonic Crystal Fibers with Multimode, Solid Core Fiber Couplers through Arc Fusion Splicing for the Miniaturization of Nonlinear Spectroscopy Sensing Devices, Fibers \textbf{6},
  77 (2018).
\bibitem{bib:thapa}
R. Thapa et al.,
  Arc fusion splicing of hollow-core photonic bandgap fibers for gas-filled fiber cells, Optics Express \textbf{14},
  9576 (2006).
\bibitem{bib:robert}
D. R. Haeupl et al.,
 Spatially resolved spectroscopy of alkali metal
vapour diffusing inside hollow-core photonic crystal
fibers, arXiv:2205.06148 (2022).
\bibitem{bib:finkelstein}
R. Finkelstein et al.,
 Fast, noise-free memory for photon synchronization at room temperature, Science Advances \textbf{4},
  eaap8598 (2018).
\bibitem{bib:lindblad}
D. A. Steck,
  Quantum and Atom Optics, available online at http://steck.us/teaching (revision
0.13.12, 25 May 2022).
\bibitem{bib:zhang}
X. Zhang et al.,
 Design of large mode area all-solid anti-resonant fiber for high-power lasers, High Power Laser Science and Engineering \textbf{9},
  e23 (2021).
\bibitem{bib:habib}
M. S. Habib et al.,
Single-mode, low loss hollow-core antiresonant fiber designs, Optics Express \textbf{27},
  3824 (2019).
\bibitem{bib:chen}
X. Chen et al.,
 High coupling efficiency technology of large core hollow-core fiber with single mode fiber, Optics Express \textbf{27},
  33135 (2019).
\end{thebibliography}
\end{document}